\documentstyle[twoside,fleqn,epsf,espcrc2,draft]{article}%
\begin{document}
\renewcommand{\refname}{\normalsize\bf References}
\newcommand{\mquad}{\!\!\!\!\!}
\title{
Correlations and fluctuations of matrix elements and cross sections
}
\author{
         Bruno Eckhardt, Imre Varga and P{\'e}ter Pollner
        \address{Fachbereich Physik, Philipps--Universit{\"a}t Marburg,\\
                 Renthof 6, D-35032 Marburg an der Lahn, Germany}
\thanks{Supported by OTKA T029813, T024136 and F024135, the Alexander
        von Humboldt foundation,the Hungarian Committee 
         for Technical Development (OMFB) and DAAD-M\"OB}
}
%
%
\begin{abstract}
\hrule
\mbox{}\\[-0.2cm]

\noindent{\bf Abstract}\\

The fluctuations and correlations of matrix elements of
cross sections are investigated in open systems that are chaotic in the
classical limit. The form of the correlation functions is discussed
within a statistical analysis and tested in calculations for 
a damped quantum kicked rotator.
We briefly comment on the modifications expected for 
systems with slowly decaying correlations,
a typical feature in mixed phase spaces.
\\[0.2cm]
\hrule
\end{abstract}

\maketitle

\section{Introduction}
The photo-dissociation cross section as determined by Fermi's golden
rule contains a combination of final density of states
and transitional matrix elements. The statistical properties of
this cross section in a situation of a chaotic dynamics
will thus be determined by the statistical properties of both the
density of states and the matrix elements. Moreover, if there is 
a classical underlying dynamics, both will be connected, in 
a suitable semiclassical limit, to the properties of the 
classical dynamics. Our aim here is to present a few
numerical and theoretical considerations connected to these
observations. 

We focus on the form of the correlation function
of photo-absorption cross sections. Using 
a random matrix theory approach Alhassid and Fyodorov \cite{AF}
found a correlation function that consisted of a Lorentzian and 
a derivative of a Lorentzian with respect to the width.
The first term is familiar from the analysis of Ericsson 
fluctuations in nuclear physics \cite{USLeshouches,BS}
and the second one from
the correlation function of the Wigner time delay
\cite{Chaos}. The 
form of the correlation function can now be made plausible
if within a statistical model for the cross section the
fluctuations in the density of states and in the transition
matrix elements are independent, as discussed in section 2.

A semiclassical analysis of this correlation function \cite{EFV}
shows that the relative weight of the two contributions depends
on the ratio of the fluctuations of the observable to the 
mean. In case of a single initial state, random matrix theory
fixes this ratio to universal numbers. However, in situations
an incoherent superposition of initial states contributes
to the cross section, variations in the relative weight are
possible. This is illustrated within a numerical analysis for
a damped kicked rotator in section 3.

The semiclassical connection also suggests certain modifications
in the correlation functions if the classical decay is not
purely exponential\cite{other}. In particular, in situations 
with mixed phase space an algebraic decay is expected\cite{chirikov}. 
The modifications
in the correlation functions include a slower decay
and the formation of a cusp at the origin, depending on
the exponent of the decay law. In section 4 we propose
a model for the form factor and analyze some of the consequences.

We conclude with a brief summary in section~5.

\section{Matrix element correlations within random matrix theory}

The quantity we focus on is the density of states weighted by
the matrix elements of the observable,
\begin{equation}
\rho_A(E)=\sum_{\mu} A_{\mu}\delta_{\eta}(E-E_{\mu}),
\label{ro}
\end{equation}
where the sum runs over the eigenstates of the system having
eigenvalues $E_{\mu}$. The photo-absorption cross section is
proportional to this expression if the observable $A$ contains
the projection onto the initial state and the dipole operator.
The expectation value of $\hat A$ can be
written as $A_{\mu }=\bar{A}+\delta A_{\mu}$, where
 $\bar{A}$ is the mean and $\delta A_{\mu}$ is the random fluctuation
around its mean. Statistically we assume
\begin{equation}
\langle\delta A_n\rangle=0, \qquad {\rm and} \qquad
\langle\delta A_n\delta A_m\rangle=\sigma^2_A\delta_{n,m}.
\label{stat}
\end{equation}
The function $\delta_{\eta}(x)$ is a Lorentzian function with half
width parameter $\eta$, normalized so that as $\eta\to 0$ 
it approaches a Dirac--$\delta$.
We take the same value of $\eta$ for all the eigenstates implying a uniform
damping. The mean density of states is simply
\begin{equation}
\bar{\rho}_A = \langle\rho_A(E)\rangle_E =
     \int_B \frac{dE}{B}\rho_A(E) = \frac{N{\bar A}}{B},
\label{meanrho}
\end{equation}
where $N$ is the number of levels and $B$ is the energy width of the
subset of the spectrum, over which the average $\langle\dots\rangle_E$ is
calculated.

The normalized autocorrelation function of the fluctuations of the density of
states, $\delta\rho_A(E)=\rho_A(E)-\bar{\rho}_A$ is defined as
\begin{eqnarray}
C(\varepsilon )\mquad &=&\mquad \frac
{\langle\delta\rho_A(E+\varepsilon )\delta\rho_A(E)\rangle_E}
{\bar{\rho}^2_A} \label{cf} \\
\mquad  &=&\mquad  \left (\frac{B}{N{\bar A}}\right )^2
\int_B \frac{dE}{B}\rho_A(E+\varepsilon)\rho_A(E) - 1.
\nonumber
\end{eqnarray}
Inserting the definition (\ref{ro}) in (\ref{cf}) we obtain
\begin{eqnarray}
C(\varepsilon)\mquad &=&\mquad \left(\frac{B}{N{\bar A}}\right)^2
\sum_{\mu,\nu}\big[{\bar A}^2+{\bar A}(\delta A_{\mu} + \delta A_{\nu})
\nonumber \\
& & \qquad \quad 
    + \delta A_{\mu}\delta A_{\nu}\big]g_{\eta}(\varepsilon,E_{\mu},E_{\nu})
    -1,
\label{cf1}
\end{eqnarray}
where we have inserted the shorthand notation
\begin{equation}
g_{\eta}(\varepsilon,E_{\mu},E_{\nu})=
\langle\delta_{\eta}(E-E_{\mu}+\varepsilon)\delta_{\eta}(E-E_{\nu})\rangle_E.
\label{shn}
\end{equation}
With the assumption that matrix elements and resonances are uncorrelated
\cite{physicad}, averaging over the $\delta A$'s eliminates two terms,
\begin{eqnarray}
C(\varepsilon)&=&
\left (\frac{B}{N}\right )^2\sum_{\mu,\nu}
                           g_{\eta}(\varepsilon,E_{\mu},E_{\nu})-1
\label{cf2} \\
&+&
\left (\frac{B}{N{\bar A}}\right )^2\sum_{\mu.\nu}
\langle\delta A_{\mu}\delta A_{\nu}\rangle g_{\eta}(\varepsilon,E_{\mu},E_{\nu}).
\nonumber
\end{eqnarray}
In the second term we will utilize (\ref{stat}).
Also we will split the first double sum in diagonal ($\mu=\nu$) and
non-diagonal ($\mu\neq\nu$) parts. For sufficiently small $\eta$ the
diagonal term is the autocorrelation function of Lorentzian
that also yields a Lorentzian.
\begin{eqnarray}
C(\varepsilon)&=&
\left (1+\frac{\sigma^2_A}{{\bar A}^2}\right )\frac{B^2}{N}
\langle\delta_{\eta}(E-\varepsilon)\delta_{\eta}(E)\rangle_E -1 \nonumber \\
 &+&
\left (\frac{B}{N}\right )^2\sum_{\mu\neq\nu}
g_{\eta}(\varepsilon,E_{\mu},E_{\nu}).
\label{cf3}
\end{eqnarray}
After performing the averaging in the first term and in the sum together with
the definition in (\ref{shn}), one arrives at
\begin{eqnarray}
C(\varepsilon)&=&
\Delta\frac{\sigma_A^2}{{\bar A}^2}\delta_{2\eta}(\varepsilon)-1 \nonumber \\
 &+&
\Delta\frac{1}{N}\sum_{\mu\neq\nu}
      \delta_{2\eta}[\varepsilon-(E_{\mu}-E_{\nu})],
\label{cf4}
\end{eqnarray}
where $\Delta =B/N$. In the sum one can recognize the appearance of the
two--level correlation function in the limit of $\eta\to 0$. For strongly
overlapping resonances, i.e.\ when $\eta\gg\Delta$ the above expression 
reduces to
\begin{equation}
C(\varepsilon)\propto
\left (\frac{\sigma_A^2}{{\bar A}^2}\frac{\eta}{\eta^2+\varepsilon^2} +
       \frac{1}{2\pi}\frac{\eta^2-\varepsilon^2}{(\eta^2+\varepsilon^2)^2}
\right )
\label{cf5}
\end{equation}
Thus the correlation function is characterized by two terms, a Lorentzian
and a derivative of a Lorentzian with respect to the broadening parameter
$\eta$. The weight of the first term comes from the fluctuations of the
observable $A$. In the case when $\hat A=|i\rangle\langle i|$ is a
projection on the basis state $|i\rangle$. This quantity in random
matrix theory (RMT) \cite{RMT} is $(\beta +2)/\beta$, where $\beta
=1,2,$ and 4 for the different universality classes the system belongs
to (orthogonal, unitary, and symplectic, respectively). Hence we
recover the correlation function derived in Ref.~\cite{AF} for
channels with uniform resonance width.

However, if $\bar{A}$ vanishes, the normalization by $\bar{A}$ is not
possible and the second term, which comes from the correlation
function of the density of states, cf.~(\ref{cf1}) and (\ref{shn}),
disappears and the correlation function becomes a pure
Lorentzian. These findings are thus in accordance with what has been
argued on the basis of semiclassical periodic orbit theory\cite{EFV}.

\section{Correlations in the quantum kicked rotator}

To illustrate the above calculations we calculate the correlation
function of the matrix element weighted density of states for a damped
kicked rotator.
 
We consider the statistical properties of observables that are projections 
onto a subset of the basis states
\begin{equation}
\hat A=\sum_{n\in I(m)} |n\rangle\langle n|,
\label{Adef}
\end{equation}
where $I(m)$ is a subset of size $m<N$ of the basis set. The fluctuations
of the matrix elements $\langle\mu |\hat A|\mu\rangle $ over the eigenstates
$\mu$ of the system describe the cross section fluctuations of the
excitation of the system from an initial state $I(m)$ to the final
state $\mu$ \cite{AF,FA,OA,EFV}. A possible dipole operator has been
absorbed into the definition of $|\mu\rangle$.

The model system we considered is a quantum kicked rotator with a kicking
potential
\begin{equation}
V(\phi )=k(\cos\phi - \sin 2\phi).
\end{equation}
It is known \cite{BS,other} that this model belongs to the unitary
universality class since the second term in the potential breaks the 
conjugation symmetry.

We have diagonalized the unitary one--step evolution operator $U$ at a
value of the classical kicking strength, $K$, where complete ergodicity
was expected ($K=7$~\cite{shepel}). The size of the system was fixed
to $N=201$. The matrix element weighted density of states was defined as
\begin{equation}
\rho_A(\phi )=\sum_{\mu}\langle\mu|{\hat A}|\mu\rangle
\left (1-e^{i(\phi -\phi_{\mu})-\Gamma/2}\right )^{-1}.
\end{equation}
The mean value of this observable is simply $\bar A=m/N$ where
$1\leq m\leq N$ is the number of states contributing in the projection.
\begin{figure}[ht]
\vspace{-0.3in}
\leavevmode
\epsfxsize=7cm
\epsfbox{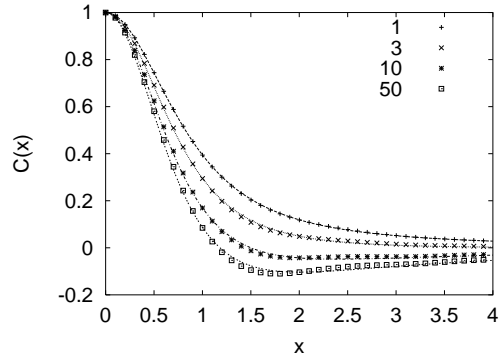}
\vspace{-0.5cm}
\caption{\label{qkr}
The correlation function $C(x)$ as a function of the eigenvalue separation
in units of mean level spacing. The data points are obtained for observables
of projections over $m$ basis states indicated in the legend. The continuous
curves are fits of the form (\protect\ref{sc-cf}) fitting $\alpha$ and
$\gamma$ independently. The width of the resonances was uniform throughout
the spectrum: $\Gamma=\Delta$.}
\vspace{-0.5cm}
\end{figure}
We have calculated the correlation function (\ref{cf}) applied for our
case
\begin{equation}
C(x)=\langle \delta\rho_A(\phi +x)\delta\rho_A(\phi) \rangle_{\phi},
\end{equation}
i.e.\ averaging $\langle\dots\rangle_{\phi}$ is done over the eigenphase
spectrum extending over $B=2\pi$. The variables $\phi$ and $x$ are measured
in units of the mean level spacing $\Delta=B/N$ therefore
$\bar\rho_A=2\pi{\bar A}/N$. The arguments of \cite{EFV} and the RMT
arguments of the previous section shows, that the correlation function
is composed of a Lorentzian and a derivative of a Lorentzian
\begin{equation}
C(x)\propto\left (\alpha\frac{\gamma}{\gamma^2+x^2}+
       \frac{1}{2\pi}\frac{\gamma^2-x^2}{(\gamma^2+x^2)^2}\right ).
\label{sc-cf}
\end{equation}
In (\ref{sc-cf}) $\alpha=\sigma^2_A/({\bar A}^2 T_H)$, with the
Heisenberg time $T_H=N$. The width $\gamma$ should come out to be the
damping $\eta$ of (\ref{cf5}), in appropriate units. 

This expression can be compared to numerical
results on the kicked rotator. In Fig.~\ref{qkr} we plot the
correlation function obtained for observables that are projections extending
over $m=1,3,10,50$ basis states. The continuous curves in the figure are
fits of the form (\ref{sc-cf}) allowing the two parameters $\alpha$ and
$\gamma$ to vary. The classical estimate for $\alpha$ is \cite{EFV,other}
\begin{equation}
\alpha=\frac{\sigma^2_A}{{\bar A}^2 N}=\frac{1-{\bar A}}{{\bar A}N}
\propto\frac{1}{m}
\label{var}
\end{equation}
The variance $\sigma^2_A$ follows its classical value for low values of
$p={\bar A}=m/N$ \cite{other}. We obtained a fitted value of
$\gamma=1.11\,\Gamma$ (instead of $\Gamma$)
independent of $m$ and $\alpha=0.284$, 0.135, 0.042, and 0.009 for
$m=1$, 3, 10, and 50, respectively. The dependence on $m$ is close to
the one predicted by (\ref{var}) at least for $m>1$.

\section{Open systems and slowly decaying correlations}

\begin{figure}[ht]
\vspace{-0.3in}
\leavevmode
\epsfxsize=7cm
\epsfbox{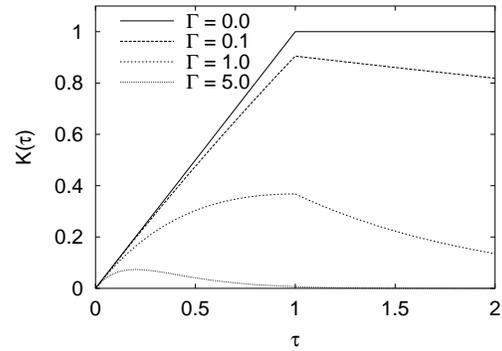}
\vspace{-0.5cm}
\caption{\label{ffexp}
The form factor for different uniform damping constants $\Gamma=0.0$, 0.1, 1.0,
and 5.0.}
\vspace{-0.5cm}
\end{figure}

\begin{figure}[ht]
\leavevmode
\epsfxsize=7cm
\epsfbox{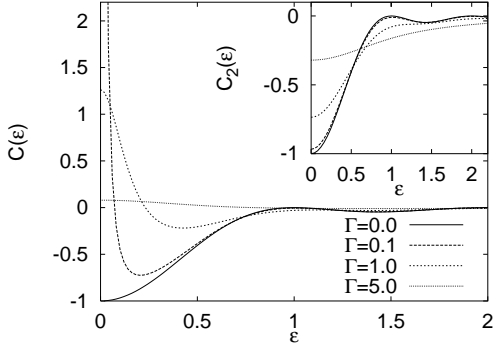}
\vspace{-0.5cm}
\caption{\label{cfexp}
The correlation function, $C(\varepsilon )$ (see Eq.~(\protect\ref{damp})), 
of an open system for different uniform damping rates $\Gamma=0.0$, 
0.1, 1.0, and 5.0. The inset shows $C_2(\varepsilon )$ 
(\protect\ref{damp-2}).}
\vspace{-0.5cm}
\end{figure}

The correlations described in the previous sections can be derived also
using the following phenomenological procedure\cite{AF}. The two--level correlation
function of a closed chaotic system, $C_0(\varepsilon)$, consists of a
Dirac--$\delta$ at the origin and a smooth function decreasing to zero for
large level separations. This is the Fourier transform of the form factor
$K_0(\tau )$. From now on for sake of simplicity we restrict ourselves to
the unitary universality class, i.e.\ we write the RMT form factor in its
standard form \cite{RMT},
$K_0(\tau )=1-b(\tau )$, with
\begin{equation}
b(\tau )=(1-|\tau |)\Theta (1-|\tau |)
\end{equation}
where $\Theta(x)$ is the step--function. Time $\tau$ is measured in units
of the Heisenberg time $T_H$, therefore energy separation $\varepsilon$,
in units of mean level spacing $\Delta$. According to RMT the
correlation function is
\begin{eqnarray}
C_0(\varepsilon )&=&\int_{-\infty}^{\infty}d\tau 
                         K_0(\tau )\cos (2\pi\varepsilon\tau) \nonumber \\
                 &=&\delta(\varepsilon )-
           \left [\frac{\sin(\pi\varepsilon)}{\pi\varepsilon}\right ]^2.
\end{eqnarray}
Uniform damping, i.e.\ opening up the system, can be introduced by
multiplying the form factor with an exponential decay.
\begin{equation}
K(\tau )=K_0(\tau )e^{-\Gamma|\tau |},
\label{ffdamp}
\end{equation}
where $\Gamma$ measures the relaxation rate in units of the Heisenberg
time. In fact $\Gamma=T_c/T_H$, where $T_c$ is the relaxation time. When
$\Gamma<1$ ($\Gamma>1$) the relaxation happens over a time scale longer 
(shorter) than $T_H$, hence produces correlations over energy scales
lower than (beyond) mean level spacing. The resulting form factor for 
different values of $\Gamma=0.0$, 0.1, 1.0 and 5.0 are plotted in
Fig.~\ref{ffexp}. The Fourier transform of (\ref{ffdamp}) results in
\begin{equation}
C(\varepsilon )=C_1(\varepsilon )+C_2(\varepsilon )
\label{damp}
\end{equation}
where
\begin{equation}
C_1(\varepsilon)=\int_{-\infty}^{\infty}d\tau
e^{-\Gamma|\tau |}\cos(2\pi\varepsilon\tau),
\end{equation}
\begin{equation}
C_2(\varepsilon)=-\int_{-1}^1 d\tau (1-|\tau |)e^{-\Gamma|\tau |}
                                 \cos(2\pi\varepsilon\tau).
\label{damp-2}
\end{equation}
The first term $C_1(\varepsilon)$ is a Lorentzian from the broadening
of the $\delta$-function and the second term, the convolution of the
Lorentzian with the two-level cluster function, leads, in the limit of
large $\Gamma$, to the derivative of the Lorentzian.
The corresponding
correlation functions $C(\varepsilon)$ are plotted in Fig.~\ref{cfexp}.

These considerations led us to extend the above approach in another direction.
Let us assume that the quantum return probability, the form factor, is
damped slower than exponential, i.e.\ in an algebraic fashion. The
correlation function in this case will contain signatures of the slowing
down of the classical dynamics, a behavior that is expected to be prominent
in systems with mixed phase space \cite{HLK}.

Such an algebraic damping may be introduced in the phenomenological ansatz
\begin{equation}
K(\tau )=K_0(\tau )\left (1+c\tau\right )^{-a},
\label{alg1}
\end{equation}
where $c=T_H/T_c$ is the ratio of the decorrelation time $T_c$ compared
to the Heisenberg time $T_H$. As in the case of exponential 
damping, for $c<1$ ($c>1$) the slow decorrelation of classical trajectories 
due to the presence of the hierarchy of stable islands \cite{HLK,rk} occur 
on a time scale that is longer (shorter) than the Heisenberg time that 
produces correlations over energy scales lower than (beyond) mean level 
spacing.
\begin{figure}[ht]
\vspace{-0.3in}
\leavevmode
\epsfxsize=7cm
\epsfbox{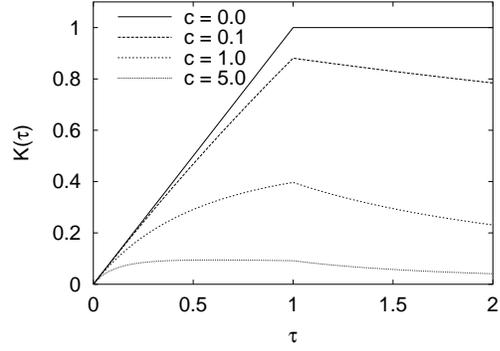}
\vspace{-0.5cm}
\caption{\label{ffpl4_3}
The form factor with algebraic damping ($a=4/3$) for different ratios of
the Heisenberg time $T_H$ compared to the decorrelation time $T_c$,
$c=T_H/T_c=0.0$, 0.1, 1.0, and 5.0.}
\vspace{-0.5cm}
\end{figure}
\begin{figure}[ht]
\leavevmode
\epsfxsize=7cm
\epsfbox{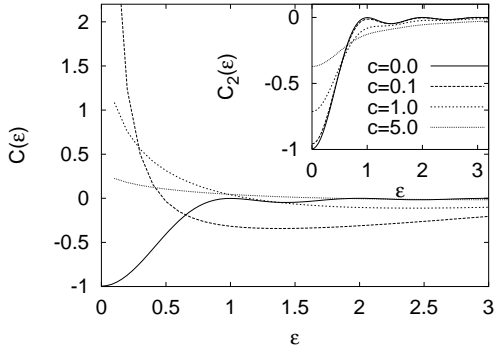}
\vspace{-0.5cm}
\caption{\label{pl-4_3}
The correlation function, $C(\varepsilon )$, of a system containing algebraic
decorrelation with exponent $a=4/3$ (see Eq.~(\protect\ref{pow}). 
The curves stand for different ratios of the Heisenberg time $T_H$ 
compared to the decorrelation time $T_c$, $c=T_H/T_c$, with $c=0.0$, 
0.1, 1.0, and 5.0. The inset shows $C_2(\varepsilon )$~(\protect\ref{pow-2}).}
\vspace{-0.5cm}
\end{figure}
By fixing one of the parameters $a$ and $c$ the variation of the other results
in changes in different parts of the correlation function. Similarly as 
in (\ref{damp}) we find
\begin{equation}
C(\varepsilon )=C_1(\varepsilon )+C_2(\varepsilon )
\label{pow}
\end{equation}
where
\begin{eqnarray}
C_1(\varepsilon)&=&\int_{-\infty}^{\infty}d\tau\frac{\cos(2\pi\varepsilon\tau)}
{(1+c|\tau |)^a}, \\
C_2(\varepsilon)&=&-\int_{-1}^1 d\tau\frac{1-|\tau |}{(1+c|\tau |)^a}
\cos(2\pi\varepsilon\tau).
\label{pow-2}
\end{eqnarray}
We have plotted the correlation function for the case when $a=4/3$. This
type of correlations can be expected to occur in systems with mixed
phase space\cite{rk}. In Fig.~\ref{pl-4_3} the curves demonstrate that  
a low value of $c$ results in only a slight modification of the correlations 
while large $c$, i.e.\ when $T_c<T_H$ the algebraic decorrelation results 
in changes both in the $\varepsilon<1$ and $\varepsilon>1$ regimes. This 
is in contrast to previous expectations \cite{rk} that the dynamics occuring 
over time scales up to $T_H$ should show up in the correlation function on 
energy scales beyond mean level spacing only.

\begin{figure}[ht]
\vspace{-0.3in}
\leavevmode
\epsfxsize=7cm
\epsfbox{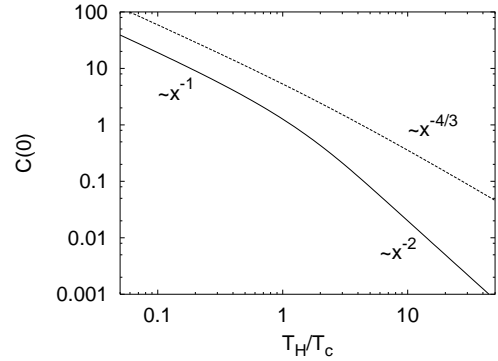}
\vspace{-0.5cm}
\caption{\label{c0fig}
The value of $C(\varepsilon )$ at $\varepsilon=0$ for exponential 
(continuous) and power law damping (dashed curve). The ratio $T_H/T_c$ is 
$\Gamma$ for exponential and $c$ for power law damping. The power law curve
was obtained for the exponent of $a=4/3$.}
\vspace{-0.5cm}
\end{figure}

To show the similarities and differences between correlation functions
obtained for exponential and power law damping we plot the value at
$\varepsilon=0$, i.e.\ the variance of the fluctuating observable under
investigation. In Fig.~\ref{c0fig} we can see that for $T_c>T_H$ the
two types of damping functions yield the same correlation time dependence,
$\propto (T_H/T_c)^{-1}$. In the case of $T_c<T_H$ the exponential 
damping produces a $(T_H/T_c)^{-2}$ dependence and the power law version a
$(T_H/T_c)^{-a}$ (in the case shown $a=4/3$).

\begin{figure}[ht]
\vspace{-0.3in}
\leavevmode
\epsfxsize=7cm
\epsfbox{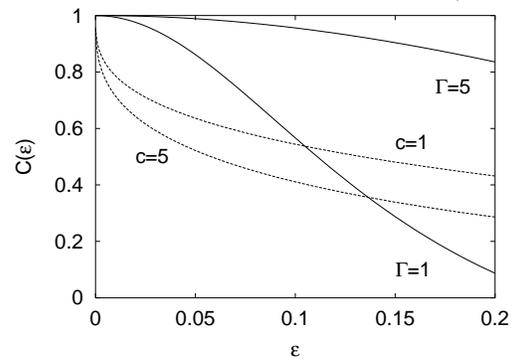}
\vspace{-0.5cm}
\caption{\label{cuspfig}
The correlation function for low--$\varepsilon$ for exponential 
(continuous) and power law damping (dashed curve) for two different values
of the decorrelation time $T_c/T_H=1$, and 5. The power law curve
was obtained for the exponent of $a=4/3$.}
\vspace{-0.5cm}
\end{figure}

An even more striking difference between the fast and slow decorrelations
shows up in the low--$\varepsilon$ behavior of the correlation function. As
pointed out already by Lai {\it et al.} \cite{Lai} non-hyperbolic systems
produce a cusp in the correlation function
\begin{equation}
C(\varepsilon )\sim C(0)-C_a\varepsilon^{a-1}
\label{cusp}
\end{equation}
where $C_a=b_ac^{-(a-1)}$ is a positive constant and $a$ is the exponent 
of the classical return probability $P(t)\sim t^{-a}$. This behavior is 
to be contrasted with the hyperbolic case when the low--$\varepsilon$ 
behavior is expected to be
\begin{equation}
C(\varepsilon )\sim C(0)-C_2\varepsilon^2,
\label{nocusp}
\end{equation}
where $C_2\sim\Gamma^{-2}$ is a constant depending on the exponential 
relaxation rate. In Fig.~\ref{cuspfig} we show the cusp at
low--$\varepsilon$ in the case of power law decorrelation  
and the parabolic behavior for exponential damping. The curves represent especially the
cases when $T_H/T_c\geq 1$, i.e.\ when the classical correlations appear at
time scales below the Heisenberg time. In this case one expects deviations
at $\varepsilon\geq 1$. The cusp at low--$\varepsilon$, however, is 
present for any value of $c$. 

We would like to emphasize that the apparent contradiction between the 
behavior of the correlation function for power law decorrelation (\ref{cusp})
and that of refs.~\cite{HLK,rk} resides probably in the difference of the
classical return probability functions considered. We took the modification
of the RMT form factor (\ref{alg1}) while in refs.~\cite{HLK,rk} the
classical return probability of $P(t)\sim t^z$ was used.

\section{Final remarks}
The variations in the correlation functions for the different
situations are not very large and perhaps difficult to detect.
This seems to apply in particular to cases with an algebraic
decay, especially in view of the fact that the exponents are not
universal and might be clouded by a distribution of algebraic
decay laws. The modifications due to variations in matrix elements
and thus a stronger emphasis of the derivative part of 
the correlation function could perhaps be achieved in 
incoherent superpositions
of cross sections from different initial states. Analysis 
of experimental data in this direction seems worthwhile.

\end{document}